\documentclass[aps,prl,preprint,groupedaddress,linenumbers]{revtex4}
 \usepackage{graphicx}
 \usepackage{dcolumn}
 \usepackage{bm}
 \usepackage{amsmath}
 \usepackage{color}
 \usepackage{amssymb}
 \usepackage{mathrsfs}
 \usepackage{MnSymbol}%
 \usepackage{wasysym}%
 \usepackage{gensymb}
 \usepackage{framed,color}
 \usepackage{enumitem}

\begin{document}



\title{Signature of chiral superconducting order parameter evidenced in mesoscopic superconductors}


\author{Xiaoying Xu$^{1}$, Wei Qin$^{2}$, Yuelin Shen$^{3}$, Zixuan Huang$^{3}$, Zhuoya Zhou$^{3}$, Zirao Wang$^{3}$, Yufan Li$^{3}$}
\affiliation{$^1$Quantum Science Center of Guangdong-HongKong-Macau Greater Bay Area, Shenzhen, Guangdong, China}
\affiliation{$^2$CAS Key Laboratory of Strongly-Coupled Quantum Matter Physics, and Department of Physics, University of Science and Technology of China, Hefei, Anhui 230026, China}
\affiliation{$^3$Department of Physics, The Chinese University of Hong Kong, Shatin, Hong Kong SAR, China}


\baselineskip24pt

\begin{abstract}
	Chiral superconductivity is a novel superconducting phase characterized by order parameters that break the time-reversal symmetry, endowing the state with a definite handedness. 
    Unlike conventional superconductors, the Cooper pairs in a chiral superconductor carry nonzero orbital angular momentum. 
    Through coupling with an external magnetic field, the finite angular momentum of the Cooper pair modulates the temperature-magnetic field phase boundary in a distinctive way, which could serve as an experimental signature of the chiral superconducting state. 
    Here we demonstrate that the chiral signature can be detected in mesoscopic superconducting rings of $\beta$-Bi$_2$Pd, manifesting as a \textit{linear-in-field} modulation of the critical temperature in the Little-Parks effect. 
    Our findings establish a new experimental method for detecting the chiral superconductivity.

\end{abstract}

\maketitle


\begin{text}

Chiral superconductivity is an unconventional superconducting state in which the Cooper pair wavefunction breaks time-reversal symmetry and carries finite orbital angular momentum \cite{Kallin2016}. 
In the momentum space, this manifests itself as a multi-component complex superconducting gap function, with the phase winding clockwise or counterclockwise around the Fermi surface. Typical examples of chiral pairing states include $p \pm i p$ and $d \pm i d$, which are intricately linked to topological superconductivity, giving rise to various exotic phenomena, such as edge currents and half-quantum vortices \cite{Volovik1996,Volovik2000,chung_stability_2007}. These characteristics position chiral superconductor as a compelling platform for the realization of Majorana fermions and non-Abelian braiding statistics \cite{DasSarma2006}.

Experimental efforts to identify chiral superconductors have primarily focused on signatures of time-reversal symmetry breaking (TRSB). 
To date, the results remain far from conclusive. 
While evidences of TRSB from muon spin relaxation and polar Kerr effect are reported in candidate materials such as Sr$_2$RuO$_4$, URu$_2$Si$_2$, UPt$_3$, and UTe$_2$ \cite{Luke1998,Xia2006,Schemm2015,Luke1993,Schemm2014,Hayes2021}, 
the anticipated chiral edge current could not be observed in scanning SQUID and scanning Hall probe experiments \cite{Kirtley2007,Curran2014,Iguchi2021,Iguchi2023,Wang2024}. 
In the case of Sr$_2$RuO$_4$, studies of the Josephson effect indicate that the superconducting state is time-reversal-invariant \cite{Kashiwaya2019}. 
It has been argued that the observed TRSB may arise from surface magnetism in the normal state \cite{Fittipaldi2021} or other extrinsic factors \cite{Willa2021}. 
In the case of UTe$_2$, with improved sample quality, recent experiments show that the time-reversal symmetry is preserved in the superconducting ground state \cite{Ajeesh2023,Azari2023}, implying extrinsic origins of the earlier reported TRSB signatures.

In light of these controversial results about the confirmation of chiral pairing states in superconductors, we instead focus on another defining feature of chiral pairing \textemdash ~the non-zero orbital angular momentum of the Cooper pairs. 
This finite angular momentum couples to an external magnetic field, modifying the free energy landscape and thus the temperature–magnetic field ($T$–$H$) phase boundary. 
We are particularly interested in the Little-Parks effect, where $T_c$ oscillates due to fluxoid quantization, a phenomenon historically vital for establishing the role of electron pairing in superconductivity \cite{Little1962}. 
More recent studies have employed the Little-Parks effect to identify half-integer fluxoid quantization linked to an anisotropic gap function \cite{Li2019} and to reveal odd-parity pairing symmetry \cite{xu_observation_2024,Yamaki2025} consistent with a $p$-wave pairing state in $\beta$-Bi$_2$Pd. 
Although the oscillatory nature of the Little-Parks effect is well understood, it also features a non-oscillatory background, generally exhibiting a quadratic dependence of the magnetic field, which is attributed to geometric factors of mesoscopic devices \cite{Moshchalkov1995}. 
In this work, we show that for a chiral superconducting state, the coupling between its finite angular momentum and an external field induces a \textit{linear-in-field} modulation of the $T$–$H$ phase boundary. 
This signature is observed in our mesoscopic rings of $\beta$-Bi$_2$Pd, suggesting the presence of a multi-component superconducting order parameter. 
Together with the observation of a $\pi$ phase shift in the Little-Parks oscillation \cite{Li2019,xu_observation_2024}, this finding leads to the conclusion of a chiral $p$-wave pairing state.

The superconducting state with a single-component order parameter $\psi$ in the presence of a magnetic field $\bm{H}=\bm{\nabla}\times\bm{A}$ can be described by the linearized Ginzburg-Landau (GL) equation $ [\rho \bm{D}^2 + \alpha(T)]\psi=0$, where $\rho = 1/2m$, $\bm{D} = -i\hbar\nabla-2e\bm{A}/c$, $m$ and $2e$ denote the effective mass and charge of the paired electrons, respectively. The GL coefficient $\alpha(T)$ denotes the inverse pairing susceptibility that changes sign at superconducting critical temperature \cite{Tinkham2004}. 
Applying the boundary condition of a mesoscopic ring \cite{Moshchalkov1995,devreese_superconducting_2000}, the $T$-$H$ phase boundary is given by
\begin{equation}\label{2}
\frac{T_{c}(H)-T_{c}(0)}{T_{c}(0)}=\frac{\xi_0^2}{R^2} \left[-\left(\dfrac{w}{2R}\right)^2 \left( \dfrac{\Phi}{\Phi_0} \right)^2- \left(n-\dfrac{\Phi}{\Phi_0}\right)^2\right],
\end{equation} 
where $R$ is the radius of the ring, $\xi_0$ is the zero-temperature coherence length, $w$ denotes the line width of the device, integer $n$ is the orbital quantum number, $\Phi_0$ is the flux quantum, and $\Phi$ is the magnetic flux threading through the ring. 
The second term on the right side of Eq.~(\ref{2}) suggests that $T_{c}$ oscillates as a function of the applied magnetic flux with a period of $\Phi_0$. 
This is the Little-Parks oscillation as a result of fluxoid quantization in the superconducting ring \cite{Little1962}. 
The first term, on the other hand, is an aperiodic background quadratically dependent on the applied field \cite{Moshchalkov1995}. 
In Fig.~1(a) we plot the $T$–$H$ phase boundary of a mesoscopic ring based on Eq.~(\ref{2}) with $R=450$~nm, and $w=100$~nm. 
It is in good agreement with the experimental data obtained in a polycrystalline Nb ring, as shown in Figs.~2(a) and 2(c). 
The experimental data can be well fitted by Eq.~(\ref{2}), yielding $\xi_0 \simeq$15~nm and $w\simeq$ 70~nm. 

For the superconducting state characterized by two degenerate order parameters $\bm{\eta} = (\eta_x,\eta_y)^{\text{T}}$, the $T$–$H$ boundary of a mesoscopic ring may differ significantly \cite{almoalem_observation_2024}. The corresponding linearized GL equation can be expressed in matrix form as \cite{Supp_2025}
\begin{equation}
\begin{pmatrix}
 \rho_L\bm{D}^2 & i\rho_T eH \\
-i\rho_T eH &  \rho_L\bm{D}^2
\end{pmatrix}
\begin{pmatrix}
\eta_x \\
\eta_y
\end{pmatrix}
=-\alpha(T)
\begin{pmatrix}
\eta_x \\
\eta_y
\end{pmatrix}
\label{eq:GLmatrix}
\end{equation}
where $\rho_L$ and $\rho_T$ denote the longitudinal and transverse superfluid stiffness, respectively. By rotating the two-component order parameter into the chiral basis, defined as $\eta_{\pm} = (\eta_x \pm i \eta_y )/\sqrt{2}$, Eq.~(\ref{eq:GLmatrix}) is reduced to 
$[\rho_L\bm{D}^2 + \alpha(T) \pm \rho_T e H ]\eta_{\pm} = 0$. 
This suggests that the presence of an out-of-plane magnetic field favors chiral superconducting states via reducing free energy. 
In the thin-wall limit, the $T$–$H$ boundary for a ring structure is determined by solving Eq.~(\ref{eq:GLmatrix}), yielding
\begin{equation}
\label{eq:clp}
\dfrac{T_{C}(H)-T_{C}(0)}{T_{C}(0)}=\dfrac{\xi_0^2}{R^2} \left[-\left(\dfrac{w}{2R}\right)^2 \left(\dfrac{\Phi}{\Phi_0}\right)^2+\frac{\rho_T}{\rho_L} \left( \dfrac{|\Phi|}{\Phi_0}\right) -\left(n-\dfrac{\Phi}{\Phi_0}\right)^2 \right].
\end{equation} 
The magnetic field favors a particular chiral superconducting state (either left-handed or right-handed), resulting in a \textit{linear-in-field} (the second term in the square bracket of Eq.~(\ref{eq:clp})) correction to the $T-H$ boundary.
The derivation process only involves the Ginzburg-Landau equation, with all the details provided in the Supplementary Material \cite{Supp_2025}.
According to Eq.~(\ref{eq:clp}), the Little-Parks effect in a superconductor with two degenerate order parameters exhibits a non-oscillatory background given by $c_1 (\Phi/\Phi_0)-c_2(\Phi/\Phi_0 )^2 $, where $c_1 =  (\rho_T/\rho_L)(\xi_0/R)^2$ and $c_2 = (w\xi_0/2R^2)^2$. 
This exotic linear component, arising from the coupling between magnetic field and the chiral order parameter, renders the $T$-$H$ phase boundary in Fig.~1(b) strikingly distinct from the conventional Little-Parks background shown in Fig.~1(a).
This contrast provides compelling, yet unexplored, experimental evidence of chiral superconductivity associated with the sample's topology.

In Fig.~2, we demonstrate the experimental data of two superconducting ring devices with identical design geometries but composed of different materials: polycrystalline Nb in Fig.~2(a) and polycrystalline $\beta$-Bi$_2$Pd in Fig.~2(b). 
The oscillation of $T_c$ observed in the Nb ring is the conventional Little-Parks effect that is well described by Eq.~(1). 
A finite magnetic field is generally expected to suppress the superconducting state. 
Therefore, the $T_c$ maximum is always observed in the absence of magnetic field. In contrast, two distinct features emerge for 
$\beta$-Bi$_2$Pd. 
First, $T_c$ reaches local minima rather than maxima at integer values of $\Phi_0$, suggesting that the flux quantum number $n$ in the oscillatory term $(n-\Phi/\Phi_0)^2$ takes half-integer numbers. 
This phenomenon is known as half-integer fluxoid quantization, resulting from unconventional pairing state. In particular, the sign change in the gap function at certain crystalline grain boundaries gives rise to a $\pi$ phase shift, which alters the typical integer quantization condition \cite{Geshkenbein1987,Li2019}. 
In $\beta$-Bi$_2$Pd, it has been shown that the half-integer quantization originates from the odd-parity pairing symmetry \cite{xu_observation_2024,Yamaki2025}. 

The second feature is even more striking: the $T_c$ 
identified from the non-oscillatory background exhibits a minimum at zero magnetic field. 
The $T_c$ maxima, in contrast, occur at finite magnetic fields, indicating that the superconductivity is enhanced in the presence of a finite magnetic field. 
This clearly deviates from the parabolic $-(\Phi/\Phi_0 )^2$ background of the conventional Little-Parks effect as described by Eq.~(1). 
Instead, we find that this feature can be captured by Eq.~(3), which includes an additional \textit{linear-in-field} term $(\Phi/\Phi_0)$. 
As shown in Fig.~2(d), a fitting curve using Eq.~(3) yields good agreement with experimental data, with fitting parameters $c_1=$1.82$\times 10^{-3}$, $c_2=$1.77$\times 10^{-4}$. 
We also obtain $\xi_0=$58.2~nm, $\rho_T/\rho_L$=0.109. 

The parabolic $c_2$ term is inversely proportional to the dimensionality of the ring, following $c_2 \propto \frac{1}{R^4}$. 
In Fig.~3(a), we plot the $T$–$H$ phase boundaries for various ring radii using Eq.~(1). 
The linear $c_1$ term also decreases as the ring size increases, exhibiting a $c_1 \propto \frac{1}{R^2}$ dependence. 
The presence of the linear component results in a dimension-dependent modification to the conventional parabolic background of the $T$–$H$ phase boundary. 
Fig.~3(b) plots the Little-Parks effect as described by Eq.~(3), using the same parameters obtained from the fitting of the $\beta$-Bi$_2$Pd ring in Fig.~2(d). 
The effect induced by the linear component is prominent for large $R$, where the $T$–$H$ boundary significantly deviate from the conventional $T$–$H$ boundary baseline with the same ring sizes (see Fig.~3(a)). 
This chiral signature becomes overwhelmed when the ring size reduces, and the conventional parabolic background is largely restored. 

We note that Eq.~(3) is derived in the vicinity of superconducting transition and is based on the doubly-degenerate chiral order parameters $\eta_{\pm}$, representing a helical state. 
In this context, an external magnetic field can energetically bias the system toward either $\eta_+$ or $\eta_-$ state.
Since reversing the direction of magnetic field also reverses the free energy landscape for $\eta_+$ and $\eta_-$, the $T$–$H$ phase boundary remains symmetric with respect to the zero magnetic field. 
However, if the normal state already exhibits TRSB, the system may favor a non-degenerate chiral superconducting state even in the vicinity of the superconducting transition. To address this situation, we may introduce a phenomenological dimensionless term $\delta$ into the GL function to lift the degeneracy between the two chiral states, yielding the following modified field dependence of $T_c$  
\begin{equation}
\dfrac{T_{C}(H)-T_{C}(0)}{T_{C}(0)} =\text{max} \left\{ \dfrac{\xi_0^2}{R^2} \left[-\left(\dfrac{w}{2R}\right)^2 \left(\dfrac{\Phi}{\Phi_0}\right)^2 \pm \frac{\rho_T}{\rho_L} \left( \dfrac{\Phi}{\Phi_0}\right)  -\left(n-\dfrac{\Phi}{\Phi_0}\right)^2 \right]\mp \delta \right\}.
\end{equation} 
The above equation introduces asymmetry to the $T$-$H$ phase boundary for opposite field polarizations, as depicted in Fig.~3(c).

We carry out a systematic experimental study varying the sizes of superconducting $\beta$-Bi$_2$Pd rings. 
The Little-Parks effect of three ring devices with different radii $R$ are shown in Fig.~4(a). 
The chiral signature of a linear $c_1$ component is evidently present, as the baseline distinctively deviates from a quadratic field dependence. 
It also confirms the dimensional dependence as predicted by Eq.~(3). 
The chiral signature is suppressed in rings with smaller radii, consistent with the behavior depicted in Fig.~3(b). 
In addition, there is no discernible asymmetry with respect to the field polarization predicted by Eq.~(4), as depicted in Fig.~3(c). 
All the three experimental curves are well fitted by Eq.~(3), with the results presented in Fig.~4(b). 
In these fittings, we use the same material parameters $\xi_0=$58.2~nm and $\rho_T/\rho_L$=0.109, which are obtained from the fitting shown in Fig.~2(d) for the ring device with $R = 450$~nm \cite{SquareRingR}. 
$w$ is the only free parameter and the best fitting results are obtained with $w$ falls in a range between 90~nm and 130~nm. 

We reach the following conclusions about the superconducting state of $\beta$-Bi$_2$Pd, compelled by the experimental observations. 
The superconducting order parameter is unconventional and multi-component. 
The fact that it could couple with an external magnetic field and modify the free energy landscape, excludes the nematic order and points to the chiral order. 
It also indicates that the TRS is not broken in the vicinity of superconducting transition, therefore the order parameter contains doubly-degenerate chiral components. 
The $\eta \pm i\eta$ order parameter is consistent with all the above criteria. 
Employing the linearized GL equation, our analysis shows that the $\eta \pm i\eta$ order parameter can explain all the key features of the $T$–$H$ phase boundary observed in mesoscopic $\beta$-Bi$_2$Pd rings. 
Taking into account the odd-parity symmetry previously reported in the system \cite{xu_observation_2024}, we conclude that the $p \pm ip$ pairing state is the most plausible option. 

Historically, the Little-Parks effect was instrumental for establishing electron pairing and the fluxoid quantization \cite{Little1962}. 
The key signature is the $\Phi_0$-periodic oscillation of $T_c$. 
Our findings show that the Little-Parks effect of unconventional superconductors could reveal far richer information than previously recognized. 
The anisotropic nature of the pairing state may facilitate a $\pi$ phase shift to the quantum oscillation of $T_c$, manifesting half-integer fluxoid quantization. 
There is also a non-oscillatory component that makes up the $T$-$H$ phase boundary in conjuncture with the Little-Parks oscillations. 
The ring geometry and finite sizes give rise to a quadratic term with respect to the applied magnetic field. 
The coupling between the magnetic field and the finite angular momentum of a multi-component, chiral order parameter induces an additional \textit{linear-in-field} term, which exhibits a different dependence on the dimensionality of the ring device. 
The competition between the two terms introduces a non-monotonic feature to the aperiodic background of the Little-Parks effect, which becomes prominent at certain characteristic length scales. 
We demonstrate that such a chiral signature in $\beta$-Bi$_2$Pd points to a helical pairing order $p \pm ip$ near the superconducting transition. 
Similar effect has also been reported in 4Hb-TaS$_2$ \cite{almoalem_observation_2024}, where an applied in-plane magnetic field is required. 
The Little-Parks effect, with both its oscillatory and non-oscillatory features, proves to be a powerful experimental probe for the superconducting pairing symmetry. 
For the case of chiral superconductivity, it offers an alternative path by passing the challenges involved in the detection of the TRSB. It is also capable of resolving the helical order comprising degenerate chiral order parameters. 
Applying this new method to the full roster of candidate chiral superconductors \cite{wang_evidence_2020,jiang_unconventional_2021,qiu_emergent_2023,le_superconducting_2024,Han2025} would mark an exciting next step.

\end{text}

\clearpage

\begin{figure}
	\centering
	\includegraphics[width=16cm]{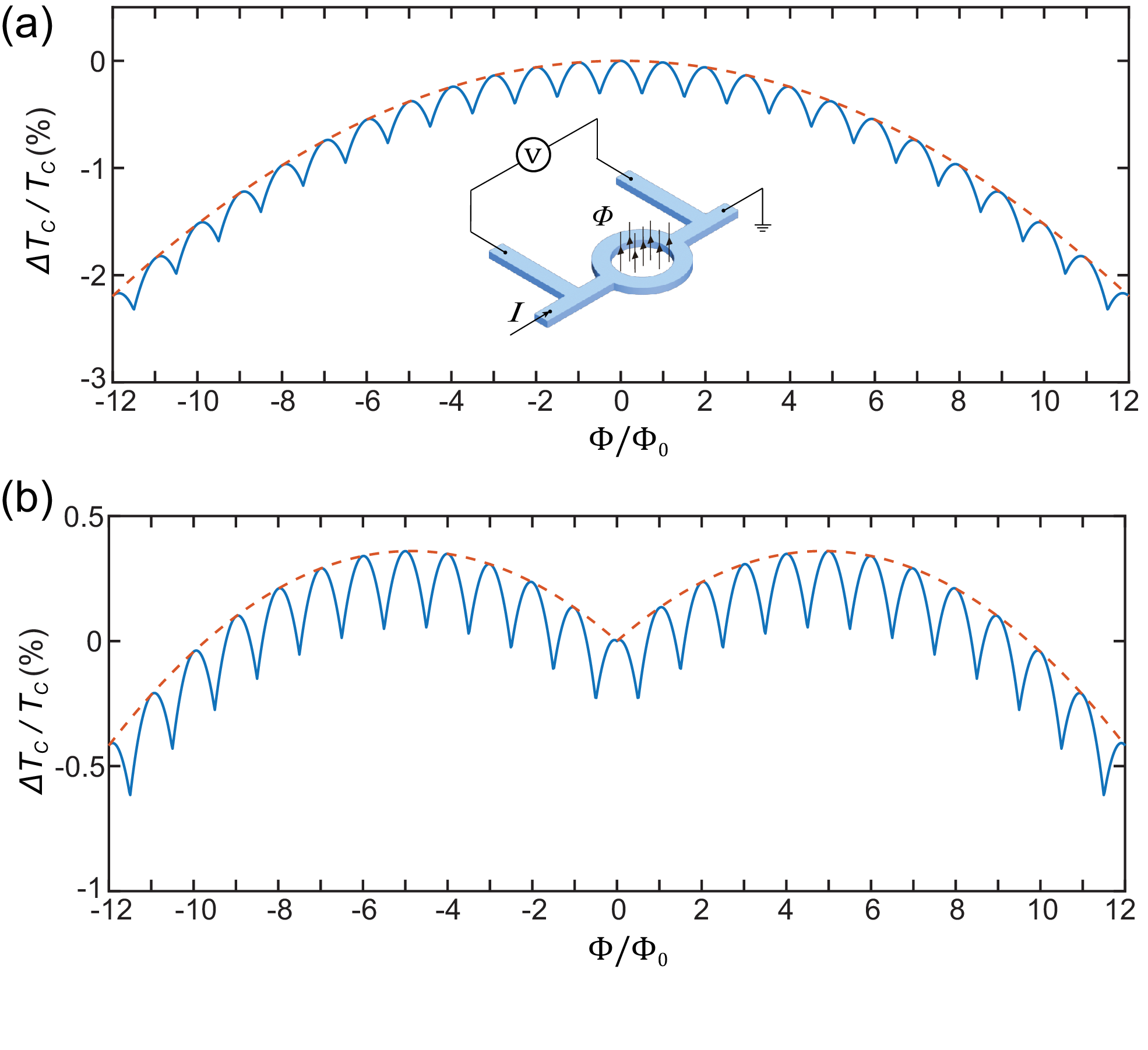}
	\caption{ The $T$-$H$ phase boundary of a mesoscopic ring device. The blue curves shows the Little-Parks effect predicted by the Ginzburg-Landau equation for a superconductor with (a) a single-component order parameter, applying Eq.~(1); and (b) two-component chiral order parameters, applying Eq.~(3). 
    The employed parameters are: $\xi_0$=50~nm, $w$=100~nm, $R$=450~nm, and $\rho_T/\rho_L$=0.12 for (b). 
    The inset is a schematic drawing of the experimental setup for the Little-Parks effect. 
    The red dashed lines are the non-oscillatory background curves of the $T$-$H$ phase boundaries. }
\end{figure}

\begin{figure}
	\centering
	\includegraphics[width=16cm]{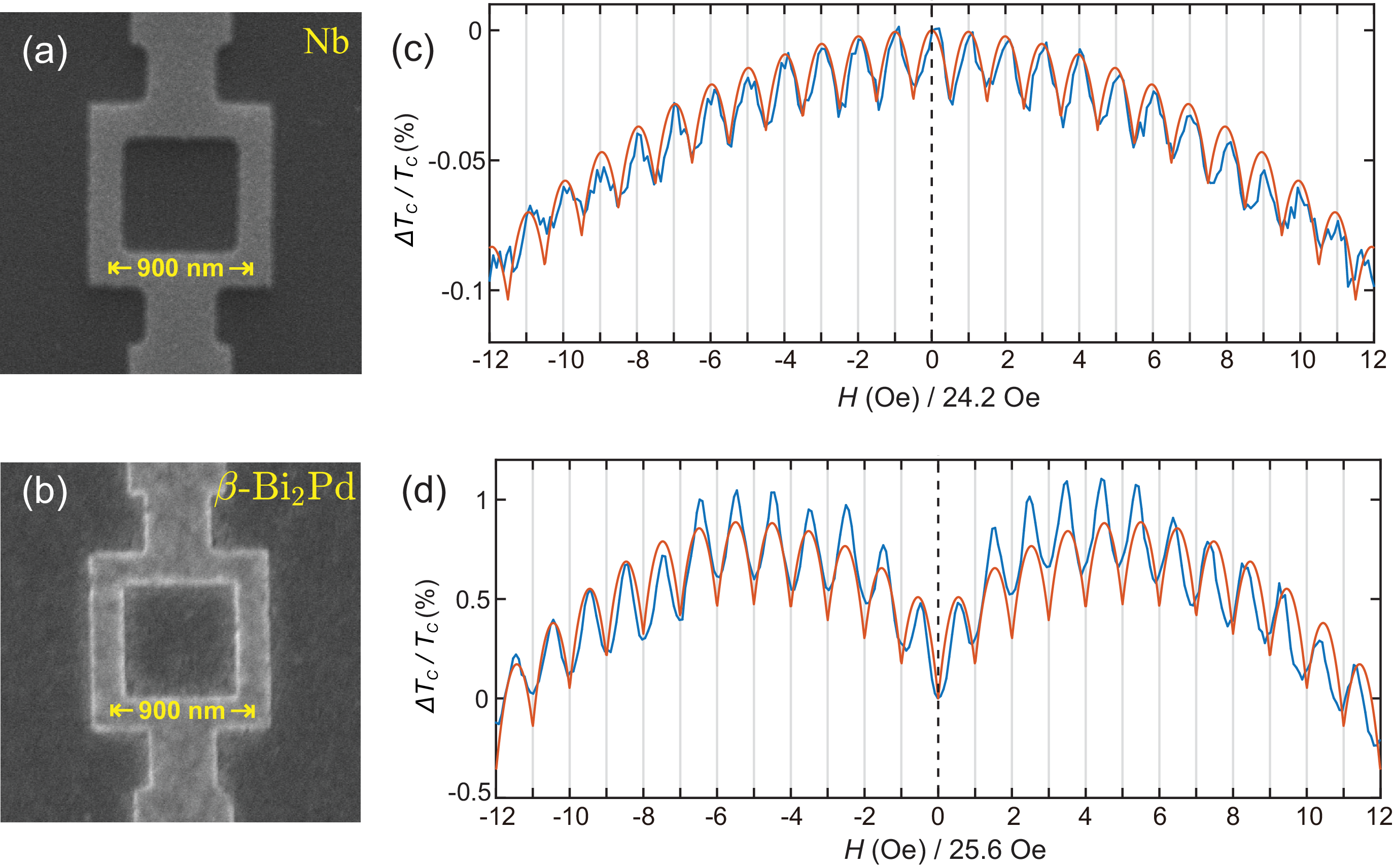}
	\caption{ The Little-Parks effect observed in Nb and $\beta$-Bi$_2$Pd rings.
    The SEM images of (a) Nb and (b) $\beta$-Bi$_2$Pd ring devices made to the same square ring geometries. 
    The side of the square, measured between the middle point of the opposing walls, is designed to be 900~nm. 
    (c) and (d) present the Little-Parks effect observed in the Nb and $\beta$-Bi$_2$Pd ring devices, respectively. 
    The blue curves are the experimental data, while the red curves are the fitting results using Eq.~(1) and Eq.~(3) for (c) and (d), respectively. }
\end{figure}

\begin{figure}
	\centering
	\includegraphics[width=6.5cm]{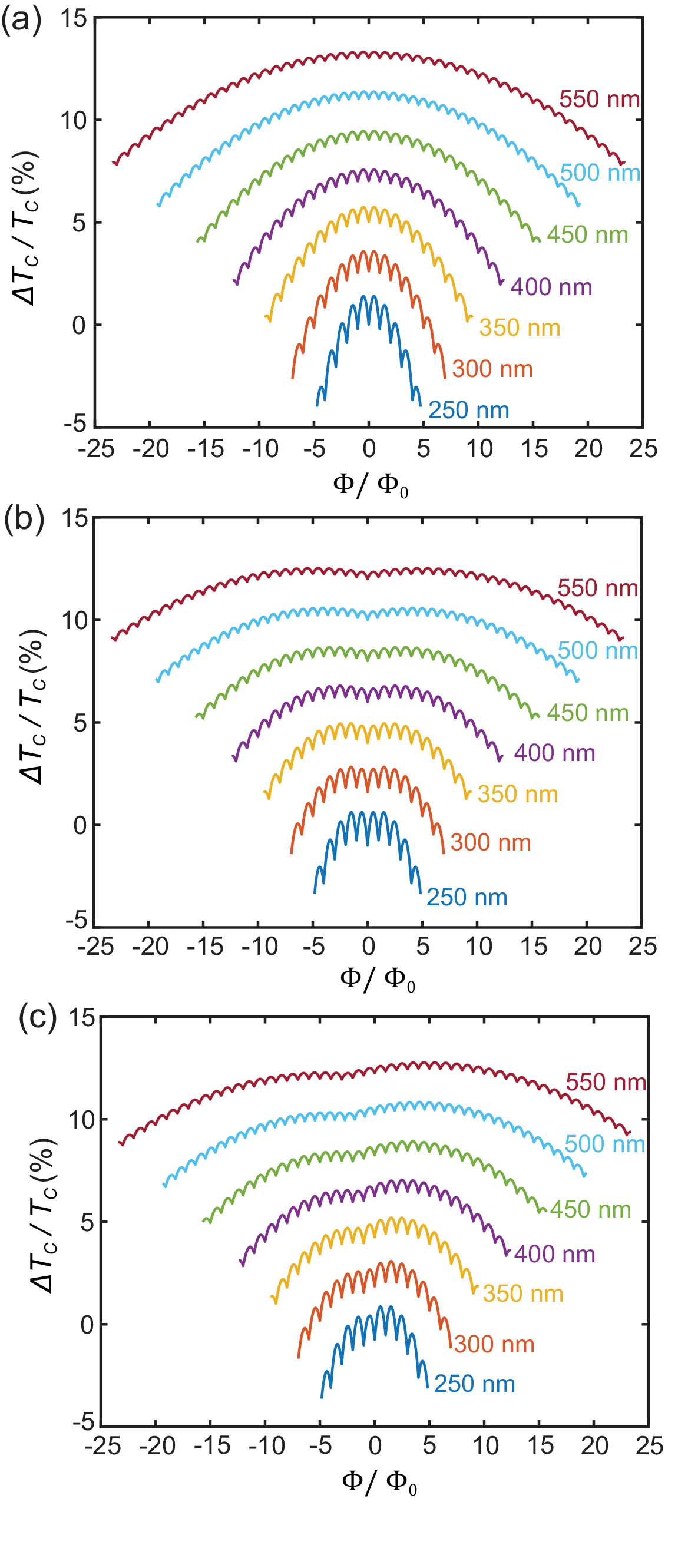}
	\caption{The $T$-$H$ phase boundaries of mesoscopic ring devices with different ring sizes.
    (a), The calculated phase boundaries of superconducting rings of a single-component order parameter following Eq. (1). 
    (b), A doubly-degenerated chiral two-component order parameter following Eq. (3).  
    (c), A TRSB chiral two-component order parameter following Eq. (4). 
    Offsets along the y-axis are applied to each curve, which originally starts from 0 when $\Phi / \Phi_0$=0.  
    The same values of $\xi_0$ and $\rho_T/\rho_L$ 
    [for (b) and (c) only] from the fitting of $\beta$-Bi$_2$Pd in Fig.~2(d) are adopted. 
    A fixed $w$=100~nm is assumed. 
    The radii of the ring devices are marked on the graph beside the corresponding curves. 
    Arbitrary shifts in the vertical axes are applied. 
    }
\end{figure}

\begin{figure}
	\centering
	\includegraphics[width=16cm]{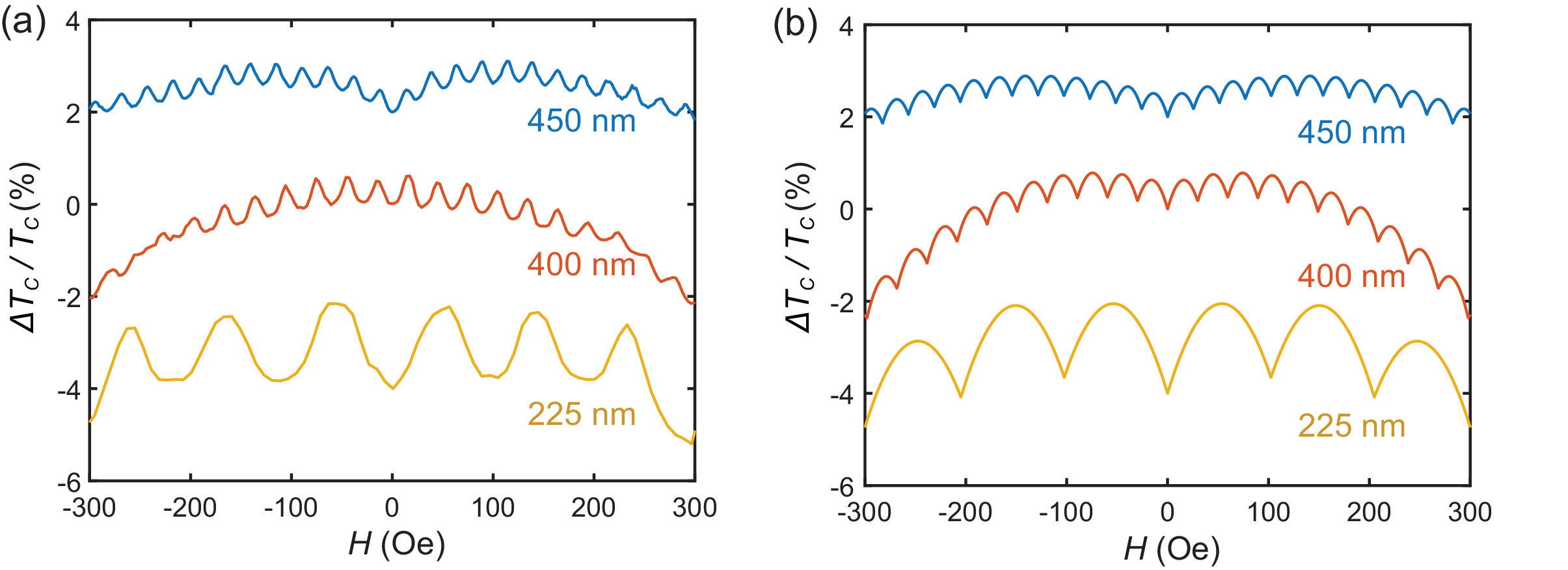}
	\caption{Little-Parks effect of three $\beta$-Bi$_2$Pd ring devices with different sizes.
    (a), The experimental data of three square rings with offsets applied along the y-axis. Half lengths of the side of the squares are 450 nm [the same device as shown in Figs. 2(c) and 2(d)], 400~nm and 225~nm, respectively. 
    (b), Calculated $T$-$H$ phase boundaries using the same ring sizes, following Eq. (3) and employing the same $\xi_0$ and $\rho_T/\rho_L$ from the fitting in Fig.~2(d). 
    Arbitrary shifts in the vertical axes are applied. 
    }
	
\end{figure}

\clearpage


\begin{thebibliography}{37}
	\expandafter\ifx\csname natexlab\endcsname\relax\def\natexlab#1{#1}\fi
	\expandafter\ifx\csname bibnamefont\endcsname\relax
	\def\bibnamefont#1{#1}\fi
	\expandafter\ifx\csname bibfnamefont\endcsname\relax
	\def\bibfnamefont#1{#1}\fi
	\expandafter\ifx\csname citenamefont\endcsname\relax
	\def\citenamefont#1{#1}\fi
	\expandafter\ifx\csname url\endcsname\relax
	\def\url#1{\texttt{#1}}\fi
	\expandafter\ifx\csname urlprefix\endcsname\relax\def\urlprefix{URL }\fi
	\providecommand{\bibinfo}[2]{#2}
	\providecommand{\eprint}[2][]{\url{#2}}
	
	\bibitem[{\citenamefont{Kallin and Berlinsky}(2016)}]{Kallin2016}
	\bibinfo{author}{\bibfnamefont{C.}~\bibnamefont{Kallin}} \bibnamefont{and}
	\bibinfo{author}{\bibfnamefont{J.}~\bibnamefont{Berlinsky}},
	\bibinfo{journal}{Rep. Prog. Phys.} \textbf{\bibinfo{volume}{79}},
	\bibinfo{pages}{054502} (\bibinfo{year}{2016}), ISSN
	\bibinfo{issn}{0034-4885},
	\urlprefix\url{http://stacks.iop.org/0034-4885/79/i=5/a=054502}.
	
	\bibitem[{\citenamefont{Volovik}(1996)}]{Volovik1996}
	\bibinfo{author}{\bibfnamefont{G.~E.} \bibnamefont{Volovik}},
	\bibinfo{journal}{Jetp Lett.} \textbf{\bibinfo{volume}{63}},
	\bibinfo{pages}{763} (\bibinfo{year}{1996}), ISSN \bibinfo{issn}{0021-3640,
		1090-6487},
	\urlprefix\url{https://link.springer.com/article/10.1134/1.566979}.
	
	\bibitem[{\citenamefont{Volovik}(2000)}]{Volovik2000}
	\bibinfo{author}{\bibfnamefont{G.~E.} \bibnamefont{Volovik}},
	\bibinfo{journal}{PNAS} \textbf{\bibinfo{volume}{97}}, \bibinfo{pages}{2431}
	(\bibinfo{year}{2000}), ISSN \bibinfo{issn}{0027-8424, 1091-6490},
	\urlprefix\url{http://www.pnas.org/content/97/6/2431}.
	
	\bibitem[{\citenamefont{Chung et~al.}(2007)\citenamefont{Chung, Bluhm, and
			Kim}}]{chung_stability_2007}
	\bibinfo{author}{\bibfnamefont{S.~B.} \bibnamefont{Chung}},
	\bibinfo{author}{\bibfnamefont{H.}~\bibnamefont{Bluhm}}, \bibnamefont{and}
	\bibinfo{author}{\bibfnamefont{E.-A.} \bibnamefont{Kim}},
	\bibinfo{journal}{Physical Review Letters} \textbf{\bibinfo{volume}{99}},
	\bibinfo{pages}{197002} (\bibinfo{year}{2007}), ISSN
	\bibinfo{issn}{0031-9007, 1079-7114},
	\urlprefix\url{https://link.aps.org/doi/10.1103/PhysRevLett.99.197002}.
	
	\bibitem[{\citenamefont{Das~Sarma et~al.}(2006)\citenamefont{Das~Sarma, Nayak,
			and Tewari}}]{DasSarma2006}
	\bibinfo{author}{\bibfnamefont{S.}~\bibnamefont{Das~Sarma}},
	\bibinfo{author}{\bibfnamefont{C.}~\bibnamefont{Nayak}}, \bibnamefont{and}
	\bibinfo{author}{\bibfnamefont{S.}~\bibnamefont{Tewari}},
	\bibinfo{journal}{Phys. Rev. B} \textbf{\bibinfo{volume}{73}},
	\bibinfo{pages}{220502} (\bibinfo{year}{2006}),
	\urlprefix\url{https://link.aps.org/doi/10.1103/PhysRevB.73.220502}.
	
	\bibitem[{\citenamefont{Luke et~al.}(1998)\citenamefont{Luke, Fudamoto, Kojima,
			Larkin, Merrin, Nachumi, Uemura, Maeno, Mao, Mori et~al.}}]{Luke1998}
	\bibinfo{author}{\bibfnamefont{G.}~\bibnamefont{Luke}},
	\bibinfo{author}{\bibfnamefont{Y.}~\bibnamefont{Fudamoto}},
	\bibinfo{author}{\bibfnamefont{K.}~\bibnamefont{Kojima}},
	\bibinfo{author}{\bibfnamefont{M.}~\bibnamefont{Larkin}},
	\bibinfo{author}{\bibfnamefont{J.}~\bibnamefont{Merrin}},
	\bibinfo{author}{\bibfnamefont{B.}~\bibnamefont{Nachumi}},
	\bibinfo{author}{\bibfnamefont{Y.}~\bibnamefont{Uemura}},
	\bibinfo{author}{\bibfnamefont{Y.}~\bibnamefont{Maeno}},
	\bibinfo{author}{\bibfnamefont{Z.}~\bibnamefont{Mao}},
	\bibinfo{author}{\bibfnamefont{Y.}~\bibnamefont{Mori}}, \bibnamefont{et~al.},
	\bibinfo{journal}{Nature} \textbf{\bibinfo{volume}{394}},
	\bibinfo{pages}{558} (\bibinfo{year}{1998}).
	
	\bibitem[{\citenamefont{Xia et~al.}(2006)\citenamefont{Xia, Maeno, Beyersdorf,
			Fejer, and Kapitulnik}}]{Xia2006}
	\bibinfo{author}{\bibfnamefont{J.}~\bibnamefont{Xia}},
	\bibinfo{author}{\bibfnamefont{Y.}~\bibnamefont{Maeno}},
	\bibinfo{author}{\bibfnamefont{P.~T.} \bibnamefont{Beyersdorf}},
	\bibinfo{author}{\bibfnamefont{M.~M.} \bibnamefont{Fejer}}, \bibnamefont{and}
	\bibinfo{author}{\bibfnamefont{A.}~\bibnamefont{Kapitulnik}},
	\bibinfo{journal}{Phys. Rev. Lett.} \textbf{\bibinfo{volume}{97}},
	\bibinfo{pages}{167002} (\bibinfo{year}{2006}),
	\urlprefix\url{https://link.aps.org/doi/10.1103/PhysRevLett.97.167002}.
	
	\bibitem[{\citenamefont{Schemm et~al.}(2015)\citenamefont{Schemm, Baumbach,
			Tobash, Ronning, Bauer, and Kapitulnik}}]{Schemm2015}
	\bibinfo{author}{\bibfnamefont{E.}~\bibnamefont{Schemm}},
	\bibinfo{author}{\bibfnamefont{R.}~\bibnamefont{Baumbach}},
	\bibinfo{author}{\bibfnamefont{P.}~\bibnamefont{Tobash}},
	\bibinfo{author}{\bibfnamefont{F.}~\bibnamefont{Ronning}},
	\bibinfo{author}{\bibfnamefont{E.}~\bibnamefont{Bauer}}, \bibnamefont{and}
	\bibinfo{author}{\bibfnamefont{A.}~\bibnamefont{Kapitulnik}},
	\bibinfo{journal}{Physical Review B} \textbf{\bibinfo{volume}{91}}
	(\bibinfo{year}{2015}).
	
	\bibitem[{\citenamefont{Luke et~al.}(1993)\citenamefont{Luke, Keren, Le, Wu,
			Uemura, Bonn, Taillefer, and Garrett}}]{Luke1993}
	\bibinfo{author}{\bibfnamefont{G.}~\bibnamefont{Luke}},
	\bibinfo{author}{\bibfnamefont{A.}~\bibnamefont{Keren}},
	\bibinfo{author}{\bibfnamefont{L.}~\bibnamefont{Le}},
	\bibinfo{author}{\bibfnamefont{W.}~\bibnamefont{Wu}},
	\bibinfo{author}{\bibfnamefont{Y.}~\bibnamefont{Uemura}},
	\bibinfo{author}{\bibfnamefont{D.}~\bibnamefont{Bonn}},
	\bibinfo{author}{\bibfnamefont{L.}~\bibnamefont{Taillefer}},
	\bibnamefont{and} \bibinfo{author}{\bibfnamefont{J.}~\bibnamefont{Garrett}},
	\bibinfo{journal}{Physical Review Letters} \textbf{\bibinfo{volume}{71}},
	\bibinfo{pages}{1466} (\bibinfo{year}{1993}).
	
	\bibitem[{\citenamefont{Schemm et~al.}(2014)\citenamefont{Schemm, Gannon,
			Wishne, Halperin, and Kapitulnik}}]{Schemm2014}
	\bibinfo{author}{\bibfnamefont{E.~R.} \bibnamefont{Schemm}},
	\bibinfo{author}{\bibfnamefont{W.~J.} \bibnamefont{Gannon}},
	\bibinfo{author}{\bibfnamefont{C.~M.} \bibnamefont{Wishne}},
	\bibinfo{author}{\bibfnamefont{W.~P.} \bibnamefont{Halperin}},
	\bibnamefont{and}
	\bibinfo{author}{\bibfnamefont{A.}~\bibnamefont{Kapitulnik}},
	\bibinfo{journal}{Science} \textbf{\bibinfo{volume}{345}},
	\bibinfo{pages}{190} (\bibinfo{year}{2014}), ISSN \bibinfo{issn}{0036-8075,
		1095-9203},
	\urlprefix\url{http://science.sciencemag.org/content/345/6193/190}.
	
	\bibitem[{\citenamefont{Hayes et~al.}(2021)\citenamefont{Hayes, Wei, Metz,
			Zhang, Eo, Ran, Saha, Collini, Butch, Agterberg et~al.}}]{Hayes2021}
	\bibinfo{author}{\bibfnamefont{I.}~\bibnamefont{Hayes}},
	\bibinfo{author}{\bibfnamefont{D.}~\bibnamefont{Wei}},
	\bibinfo{author}{\bibfnamefont{T.}~\bibnamefont{Metz}},
	\bibinfo{author}{\bibfnamefont{J.}~\bibnamefont{Zhang}},
	\bibinfo{author}{\bibfnamefont{Y.}~\bibnamefont{Eo}},
	\bibinfo{author}{\bibfnamefont{S.}~\bibnamefont{Ran}},
	\bibinfo{author}{\bibfnamefont{S.}~\bibnamefont{Saha}},
	\bibinfo{author}{\bibfnamefont{J.}~\bibnamefont{Collini}},
	\bibinfo{author}{\bibfnamefont{N.}~\bibnamefont{Butch}},
	\bibinfo{author}{\bibfnamefont{D.}~\bibnamefont{Agterberg}},
	\bibnamefont{et~al.}, \bibinfo{journal}{Science}
	\textbf{\bibinfo{volume}{373}}, \bibinfo{pages}{797} (\bibinfo{year}{2021}).
	
	\bibitem[{\citenamefont{Kirtley et~al.}(2007)\citenamefont{Kirtley, Kallin,
			Hicks, Kim, Liu, Moler, Maeno, and Nelson}}]{Kirtley2007}
	\bibinfo{author}{\bibfnamefont{J.~R.} \bibnamefont{Kirtley}},
	\bibinfo{author}{\bibfnamefont{C.}~\bibnamefont{Kallin}},
	\bibinfo{author}{\bibfnamefont{C.~W.} \bibnamefont{Hicks}},
	\bibinfo{author}{\bibfnamefont{E.-A.} \bibnamefont{Kim}},
	\bibinfo{author}{\bibfnamefont{Y.}~\bibnamefont{Liu}},
	\bibinfo{author}{\bibfnamefont{K.~A.} \bibnamefont{Moler}},
	\bibinfo{author}{\bibfnamefont{Y.}~\bibnamefont{Maeno}}, \bibnamefont{and}
	\bibinfo{author}{\bibfnamefont{K.~D.} \bibnamefont{Nelson}},
	\bibinfo{journal}{Phys. Rev. B} \textbf{\bibinfo{volume}{76}},
	\bibinfo{pages}{014526} (\bibinfo{year}{2007}),
	\urlprefix\url{https://link.aps.org/doi/10.1103/PhysRevB.76.014526}.
	
	\bibitem[{\citenamefont{Curran et~al.}(2014)\citenamefont{Curran, Bending,
			Desoky, Gibbs, Lee, and Mackenzie}}]{Curran2014}
	\bibinfo{author}{\bibfnamefont{P.}~\bibnamefont{Curran}},
	\bibinfo{author}{\bibfnamefont{S.}~\bibnamefont{Bending}},
	\bibinfo{author}{\bibfnamefont{W.}~\bibnamefont{Desoky}},
	\bibinfo{author}{\bibfnamefont{A.}~\bibnamefont{Gibbs}},
	\bibinfo{author}{\bibfnamefont{S.}~\bibnamefont{Lee}}, \bibnamefont{and}
	\bibinfo{author}{\bibfnamefont{A.}~\bibnamefont{Mackenzie}},
	\bibinfo{journal}{Physical Review B} \textbf{\bibinfo{volume}{89}}
	(\bibinfo{year}{2014}).
	
	\bibitem[{\citenamefont{Iguchi et~al.}(2021)\citenamefont{Iguchi, Zhang, Bauer,
			Ronning, Kirtley, and Moler}}]{Iguchi2021}
	\bibinfo{author}{\bibfnamefont{Y.}~\bibnamefont{Iguchi}},
	\bibinfo{author}{\bibfnamefont{I.~P.} \bibnamefont{Zhang}},
	\bibinfo{author}{\bibfnamefont{E.~D.} \bibnamefont{Bauer}},
	\bibinfo{author}{\bibfnamefont{F.}~\bibnamefont{Ronning}},
	\bibinfo{author}{\bibfnamefont{J.~R.} \bibnamefont{Kirtley}},
	\bibnamefont{and} \bibinfo{author}{\bibfnamefont{K.~A.} \bibnamefont{Moler}},
	\bibinfo{journal}{Physical Review B} \textbf{\bibinfo{volume}{103}},
	\bibinfo{pages}{l220503} (\bibinfo{year}{2021}), ISSN
	\bibinfo{issn}{2469-9969}.
	
	\bibitem[{\citenamefont{Iguchi et~al.}(2023)\citenamefont{Iguchi, Man, Thomas,
			Ronning, Rosa, and Moler}}]{Iguchi2023}
	\bibinfo{author}{\bibfnamefont{Y.}~\bibnamefont{Iguchi}},
	\bibinfo{author}{\bibfnamefont{H.}~\bibnamefont{Man}},
	\bibinfo{author}{\bibfnamefont{S.}~\bibnamefont{Thomas}},
	\bibinfo{author}{\bibfnamefont{F.}~\bibnamefont{Ronning}},
	\bibinfo{author}{\bibfnamefont{P.}~\bibnamefont{Rosa}}, \bibnamefont{and}
	\bibinfo{author}{\bibfnamefont{K.}~\bibnamefont{Moler}},
	\bibinfo{journal}{Physical Review Letters} \textbf{\bibinfo{volume}{130}}
	(\bibinfo{year}{2023}).
	
	\bibitem[{\citenamefont{Wang et~al.}(2024)\citenamefont{Wang, Yao, Winyard,
			Broyles, Gould, He, Zhang, Yao, Zhu, Xiang et~al.}}]{Wang2024}
	\bibinfo{author}{\bibfnamefont{Y.~F.} \bibnamefont{Wang}},
	\bibinfo{author}{\bibfnamefont{H.~X.} \bibnamefont{Yao}},
	\bibinfo{author}{\bibfnamefont{T.}~\bibnamefont{Winyard}},
	\bibinfo{author}{\bibfnamefont{C.}~\bibnamefont{Broyles}},
	\bibinfo{author}{\bibfnamefont{S.}~\bibnamefont{Gould}},
	\bibinfo{author}{\bibfnamefont{Q.~S.} \bibnamefont{He}},
	\bibinfo{author}{\bibfnamefont{P.~H.} \bibnamefont{Zhang}},
	\bibinfo{author}{\bibfnamefont{K.~Z.} \bibnamefont{Yao}},
	\bibinfo{author}{\bibfnamefont{J.~J.} \bibnamefont{Zhu}},
	\bibinfo{author}{\bibfnamefont{B.~K.} \bibnamefont{Xiang}},
	\bibnamefont{et~al.}, \emph{\bibinfo{title}{Observation of vortex stripes in
			{{UTe}}\$\_2\$}} (\bibinfo{year}{2024}), \eprint{2408.06209}.
	
	\bibitem[{\citenamefont{Kashiwaya et~al.}(2019)\citenamefont{Kashiwaya, Saitoh,
			Kashiwaya, Koyanagi, Sato, Yada, Tanaka, and Maeno}}]{Kashiwaya2019}
	\bibinfo{author}{\bibfnamefont{S.}~\bibnamefont{Kashiwaya}},
	\bibinfo{author}{\bibfnamefont{K.}~\bibnamefont{Saitoh}},
	\bibinfo{author}{\bibfnamefont{H.}~\bibnamefont{Kashiwaya}},
	\bibinfo{author}{\bibfnamefont{M.}~\bibnamefont{Koyanagi}},
	\bibinfo{author}{\bibfnamefont{M.}~\bibnamefont{Sato}},
	\bibinfo{author}{\bibfnamefont{K.}~\bibnamefont{Yada}},
	\bibinfo{author}{\bibfnamefont{Y.}~\bibnamefont{Tanaka}}, \bibnamefont{and}
	\bibinfo{author}{\bibfnamefont{Y.}~\bibnamefont{Maeno}},
	\bibinfo{journal}{Physical Review B} \textbf{\bibinfo{volume}{100}},
	\bibinfo{pages}{094530} (\bibinfo{year}{2019}), ISSN
	\bibinfo{issn}{2469-9969}.
	
	\bibitem[{\citenamefont{Fittipaldi et~al.}(2021)\citenamefont{Fittipaldi,
			Hartmann, Mercaldo, Komori, Bjørlig, Kyung, Yasui, Miyoshi, Olde~Olthof,
			Palomares~Garcia et~al.}}]{Fittipaldi2021}
	\bibinfo{author}{\bibfnamefont{R.}~\bibnamefont{Fittipaldi}},
	\bibinfo{author}{\bibfnamefont{R.}~\bibnamefont{Hartmann}},
	\bibinfo{author}{\bibfnamefont{M.~T.} \bibnamefont{Mercaldo}},
	\bibinfo{author}{\bibfnamefont{S.}~\bibnamefont{Komori}},
	\bibinfo{author}{\bibfnamefont{A.}~\bibnamefont{Bjørlig}},
	\bibinfo{author}{\bibfnamefont{W.}~\bibnamefont{Kyung}},
	\bibinfo{author}{\bibfnamefont{Y.}~\bibnamefont{Yasui}},
	\bibinfo{author}{\bibfnamefont{T.}~\bibnamefont{Miyoshi}},
	\bibinfo{author}{\bibfnamefont{L.~A.~B.} \bibnamefont{Olde~Olthof}},
	\bibinfo{author}{\bibfnamefont{C.~M.} \bibnamefont{Palomares~Garcia}},
	\bibnamefont{et~al.}, \bibinfo{journal}{Nature Communications}
	\textbf{\bibinfo{volume}{12}} (\bibinfo{year}{2021}), ISSN
	\bibinfo{issn}{2041-1723}.
	
	\bibitem[{\citenamefont{Willa et~al.}(2021)\citenamefont{Willa, Hecker,
			Fernandes, and Schmalian}}]{Willa2021}
	\bibinfo{author}{\bibfnamefont{R.}~\bibnamefont{Willa}},
	\bibinfo{author}{\bibfnamefont{M.}~\bibnamefont{Hecker}},
	\bibinfo{author}{\bibfnamefont{R.~M.} \bibnamefont{Fernandes}},
	\bibnamefont{and}
	\bibinfo{author}{\bibfnamefont{J.}~\bibnamefont{Schmalian}},
	\bibinfo{journal}{Physical Review B} \textbf{\bibinfo{volume}{104}},
	\bibinfo{pages}{024511} (\bibinfo{year}{2021}), ISSN
	\bibinfo{issn}{2469-9969}.
	
	\bibitem[{\citenamefont{Ajeesh et~al.}(2023)\citenamefont{Ajeesh, Bordelon,
			Girod, Mishra, Ronning, Bauer, Maiorov, Thompson, Rosa, and
			Thomas}}]{Ajeesh2023}
	\bibinfo{author}{\bibfnamefont{M.}~\bibnamefont{Ajeesh}},
	\bibinfo{author}{\bibfnamefont{M.}~\bibnamefont{Bordelon}},
	\bibinfo{author}{\bibfnamefont{C.}~\bibnamefont{Girod}},
	\bibinfo{author}{\bibfnamefont{S.}~\bibnamefont{Mishra}},
	\bibinfo{author}{\bibfnamefont{F.}~\bibnamefont{Ronning}},
	\bibinfo{author}{\bibfnamefont{E.}~\bibnamefont{Bauer}},
	\bibinfo{author}{\bibfnamefont{B.}~\bibnamefont{Maiorov}},
	\bibinfo{author}{\bibfnamefont{J.}~\bibnamefont{Thompson}},
	\bibinfo{author}{\bibfnamefont{P.}~\bibnamefont{Rosa}}, \bibnamefont{and}
	\bibinfo{author}{\bibfnamefont{S.}~\bibnamefont{Thomas}},
	\bibinfo{journal}{Physical Review X} \textbf{\bibinfo{volume}{13}},
	\bibinfo{pages}{041019} (\bibinfo{year}{2023}), ISSN
	\bibinfo{issn}{2160-3308}.
	
	\bibitem[{\citenamefont{Azari et~al.}(2023)\citenamefont{Azari, Yakovlev, Rye,
			Dunsiger, Sundar, Bordelon, Thomas, Thompson, Rosa, and Sonier}}]{Azari2023}
	\bibinfo{author}{\bibfnamefont{N.}~\bibnamefont{Azari}},
	\bibinfo{author}{\bibfnamefont{M.}~\bibnamefont{Yakovlev}},
	\bibinfo{author}{\bibfnamefont{N.}~\bibnamefont{Rye}},
	\bibinfo{author}{\bibfnamefont{S.}~\bibnamefont{Dunsiger}},
	\bibinfo{author}{\bibfnamefont{S.}~\bibnamefont{Sundar}},
	\bibinfo{author}{\bibfnamefont{M.}~\bibnamefont{Bordelon}},
	\bibinfo{author}{\bibfnamefont{S.}~\bibnamefont{Thomas}},
	\bibinfo{author}{\bibfnamefont{J.}~\bibnamefont{Thompson}},
	\bibinfo{author}{\bibfnamefont{P.}~\bibnamefont{Rosa}}, \bibnamefont{and}
	\bibinfo{author}{\bibfnamefont{J.}~\bibnamefont{Sonier}},
	\bibinfo{journal}{Physical Review Letters} \textbf{\bibinfo{volume}{131}}
	(\bibinfo{year}{2023}).
	
	\bibitem[{\citenamefont{Little and Parks}(1962)}]{Little1962}
	\bibinfo{author}{\bibfnamefont{W.~A.} \bibnamefont{Little}} \bibnamefont{and}
	\bibinfo{author}{\bibfnamefont{R.~D.} \bibnamefont{Parks}},
	\bibinfo{journal}{Phys. Rev. Lett.} \textbf{\bibinfo{volume}{9}},
	\bibinfo{pages}{9} (\bibinfo{year}{1962}),
	\urlprefix\url{https://link.aps.org/doi/10.1103/PhysRevLett.9.9}.
	
	\bibitem[{\citenamefont{Li et~al.}(2019)\citenamefont{Li, Xu, Lee, Chu, and
			Chien}}]{Li2019}
	\bibinfo{author}{\bibfnamefont{Y.}~\bibnamefont{Li}},
	\bibinfo{author}{\bibfnamefont{X.}~\bibnamefont{Xu}},
	\bibinfo{author}{\bibfnamefont{M.-H.} \bibnamefont{Lee}},
	\bibinfo{author}{\bibfnamefont{M.-W.} \bibnamefont{Chu}}, \bibnamefont{and}
	\bibinfo{author}{\bibfnamefont{C.~L.} \bibnamefont{Chien}},
	\bibinfo{journal}{Science} \textbf{\bibinfo{volume}{366}},
	\bibinfo{pages}{238} (\bibinfo{year}{2019}),
	\eprint{https://www.science.org/doi/pdf/10.1126/science.aau6539},
	\urlprefix\url{https://www.science.org/doi/abs/10.1126/science.aau6539}.
	
	\bibitem[{\citenamefont{Xu et~al.}(2024)\citenamefont{Xu, Li, and
			Chien}}]{xu_observation_2024}
	\bibinfo{author}{\bibfnamefont{X.}~\bibnamefont{Xu}},
	\bibinfo{author}{\bibfnamefont{Y.}~\bibnamefont{Li}}, \bibnamefont{and}
	\bibinfo{author}{\bibfnamefont{C.}~\bibnamefont{Chien}},
	\bibinfo{journal}{Physical Review Letters} \textbf{\bibinfo{volume}{132}},
	\bibinfo{pages}{056001} (\bibinfo{year}{2024}), \bibinfo{note}{publisher:
		American Physical Society},
	\urlprefix\url{https://link.aps.org/doi/10.1103/PhysRevLett.132.056001}.
	
	\bibitem[{\citenamefont{Yamaki and Asano}(2025)}]{Yamaki2025}
	\bibinfo{author}{\bibfnamefont{K.}~\bibnamefont{Yamaki}} \bibnamefont{and}
	\bibinfo{author}{\bibfnamefont{Y.}~\bibnamefont{Asano}},
	\bibinfo{journal}{Physical Review B} \textbf{\bibinfo{volume}{111}}
	(\bibinfo{year}{2025}).
	
	\bibitem[{\citenamefont{Moshchalkov et~al.}(1995)\citenamefont{Moshchalkov,
			Gielen, Strunk, Jonckheere, Qiu, Haesendonck, and
			Bruynseraede}}]{Moshchalkov1995}
	\bibinfo{author}{\bibfnamefont{V.~V.} \bibnamefont{Moshchalkov}},
	\bibinfo{author}{\bibfnamefont{L.}~\bibnamefont{Gielen}},
	\bibinfo{author}{\bibfnamefont{C.}~\bibnamefont{Strunk}},
	\bibinfo{author}{\bibfnamefont{R.}~\bibnamefont{Jonckheere}},
	\bibinfo{author}{\bibfnamefont{X.}~\bibnamefont{Qiu}},
	\bibinfo{author}{\bibfnamefont{C.~V.} \bibnamefont{Haesendonck}},
	\bibnamefont{and}
	\bibinfo{author}{\bibfnamefont{Y.}~\bibnamefont{Bruynseraede}},
	\bibinfo{journal}{Nature} \textbf{\bibinfo{volume}{373}},
	\bibinfo{pages}{319} (\bibinfo{year}{1995}), ISSN \bibinfo{issn}{1476-4687},
	\urlprefix\url{https://doi.org/10.1038/373319a0}.
	
	\bibitem[{\citenamefont{Tinkham}(2004)}]{Tinkham2004}
	\bibinfo{author}{\bibfnamefont{M.}~\bibnamefont{Tinkham}},
	\emph{\bibinfo{title}{{Introduction to Superconductivity: Second Edition
				(Dover Books on Physics) (Vol i)}}} (\bibinfo{publisher}{Dover Publications},
	\bibinfo{year}{2004}), \bibinfo{edition}{second edition} ed., ISBN
	\bibinfo{isbn}{0486435032},
	\urlprefix\url{http://www.amazon.com/exec/obidos/redirect?tag=citeulike07-20\&path=ASIN/0486435032}.
	
	\bibitem[{\citenamefont{Devreese}(2000)}]{devreese_superconducting_2000}
	\bibinfo{author}{\bibfnamefont{J.~T.} \bibnamefont{Devreese}},
	\bibinfo{journal}{Physica C: Superconductivity}
	\textbf{\bibinfo{volume}{332}}, \bibinfo{pages}{242} (\bibinfo{year}{2000}),
	ISSN \bibinfo{issn}{0921-4534},
	\urlprefix\url{https://www.sciencedirect.com/science/article/pii/S0921453499006796}.
	
	\bibitem[{\citenamefont{Almoalem et~al.}(2024)\citenamefont{Almoalem, Feldman,
			Mangel, Shlafman, Yaish, Fischer, Moshe, Ruhman, and
			Kanigel}}]{almoalem_observation_2024}
	\bibinfo{author}{\bibfnamefont{A.}~\bibnamefont{Almoalem}},
	\bibinfo{author}{\bibfnamefont{I.}~\bibnamefont{Feldman}},
	\bibinfo{author}{\bibfnamefont{I.}~\bibnamefont{Mangel}},
	\bibinfo{author}{\bibfnamefont{M.}~\bibnamefont{Shlafman}},
	\bibinfo{author}{\bibfnamefont{Y.~E.} \bibnamefont{Yaish}},
	\bibinfo{author}{\bibfnamefont{M.~H.} \bibnamefont{Fischer}},
	\bibinfo{author}{\bibfnamefont{M.}~\bibnamefont{Moshe}},
	\bibinfo{author}{\bibfnamefont{J.}~\bibnamefont{Ruhman}}, \bibnamefont{and}
	\bibinfo{author}{\bibfnamefont{A.}~\bibnamefont{Kanigel}},
	\bibinfo{journal}{Nature Communications} \textbf{\bibinfo{volume}{15}},
	\bibinfo{pages}{4623} (\bibinfo{year}{2024}), ISSN \bibinfo{issn}{2041-1723},
	\bibinfo{note}{publisher: Nature Publishing Group},
	\urlprefix\url{https://www.nature.com/articles/s41467-024-48260-x}.
	
	\bibitem[{Sup(2025)}]{Supp_2025}
	\bibinfo{journal}{See Supplemental Material at
		URL-will-be-inserted-by-publisher for the data of the experiments.}
	(\bibinfo{year}{2025}).
	
	\bibitem[{\citenamefont{Geshkenbein et~al.}(1987)\citenamefont{Geshkenbein,
			Larkin, and Barone}}]{Geshkenbein1987}
	\bibinfo{author}{\bibfnamefont{V.~B.} \bibnamefont{Geshkenbein}},
	\bibinfo{author}{\bibfnamefont{A.~I.} \bibnamefont{Larkin}},
	\bibnamefont{and} \bibinfo{author}{\bibfnamefont{A.}~\bibnamefont{Barone}},
	\bibinfo{journal}{Physical Review B} \textbf{\bibinfo{volume}{36}},
	\bibinfo{pages}{235} (\bibinfo{year}{1987}), ISSN \bibinfo{issn}{0163-1829},
	\urlprefix\url{https://link.aps.org/doi/10.1103/PhysRevB.36.235}.
	
	\bibitem[{Squ()}]{SquareRingR}
	\bibinfo{note}{R for the square ring devices is defined as half length of the
		side of the square, measured between the mid points of the opposing sides.
		Considerations for this approximation is discussed in the Supplementary
		Materials Section IV.}
	
	\bibitem[{\citenamefont{Wang et~al.}(2020)\citenamefont{Wang, Kim, Liu,
			Cevallos, Cava, and Ong}}]{wang_evidence_2020}
	\bibinfo{author}{\bibfnamefont{W.}~\bibnamefont{Wang}},
	\bibinfo{author}{\bibfnamefont{S.}~\bibnamefont{Kim}},
	\bibinfo{author}{\bibfnamefont{M.}~\bibnamefont{Liu}},
	\bibinfo{author}{\bibfnamefont{F.~A.} \bibnamefont{Cevallos}},
	\bibinfo{author}{\bibfnamefont{R.~J.} \bibnamefont{Cava}}, \bibnamefont{and}
	\bibinfo{author}{\bibfnamefont{N.~P.} \bibnamefont{Ong}},
	\bibinfo{journal}{Science} \textbf{\bibinfo{volume}{368}},
	\bibinfo{pages}{534} (\bibinfo{year}{2020}), \bibinfo{note}{publisher:
		American Association for the Advancement of Science},
	\urlprefix\url{https://www.science.org/doi/10.1126/science.aaw9270}.
	
	\bibitem[{\citenamefont{Jiang et~al.}(2021)\citenamefont{Jiang, Yin, Denner,
			Shumiya, Ortiz, Xu, Guguchia, He, Hossain, Liu
			et~al.}}]{jiang_unconventional_2021}
	\bibinfo{author}{\bibfnamefont{Y.-X.} \bibnamefont{Jiang}},
	\bibinfo{author}{\bibfnamefont{J.-X.} \bibnamefont{Yin}},
	\bibinfo{author}{\bibfnamefont{M.~M.} \bibnamefont{Denner}},
	\bibinfo{author}{\bibfnamefont{N.}~\bibnamefont{Shumiya}},
	\bibinfo{author}{\bibfnamefont{B.~R.} \bibnamefont{Ortiz}},
	\bibinfo{author}{\bibfnamefont{G.}~\bibnamefont{Xu}},
	\bibinfo{author}{\bibfnamefont{Z.}~\bibnamefont{Guguchia}},
	\bibinfo{author}{\bibfnamefont{J.}~\bibnamefont{He}},
	\bibinfo{author}{\bibfnamefont{M.~S.} \bibnamefont{Hossain}},
	\bibinfo{author}{\bibfnamefont{X.}~\bibnamefont{Liu}}, \bibnamefont{et~al.},
	\bibinfo{journal}{Nature Materials} \textbf{\bibinfo{volume}{20}},
	\bibinfo{pages}{1353} (\bibinfo{year}{2021}), ISSN \bibinfo{issn}{1476-4660},
	\bibinfo{note}{publisher: Nature Publishing Group},
	\urlprefix\url{https://www.nature.com/articles/s41563-021-01034-y}.
	
	\bibitem[{\citenamefont{Qiu et~al.}(2023)\citenamefont{Qiu, Yang, Hu, Zhang,
			Chen, Lyu, Eckberg, Deng, Krylyuk, Davydov et~al.}}]{qiu_emergent_2023}
	\bibinfo{author}{\bibfnamefont{G.}~\bibnamefont{Qiu}},
	\bibinfo{author}{\bibfnamefont{H.-Y.} \bibnamefont{Yang}},
	\bibinfo{author}{\bibfnamefont{L.}~\bibnamefont{Hu}},
	\bibinfo{author}{\bibfnamefont{H.}~\bibnamefont{Zhang}},
	\bibinfo{author}{\bibfnamefont{C.-Y.} \bibnamefont{Chen}},
	\bibinfo{author}{\bibfnamefont{Y.}~\bibnamefont{Lyu}},
	\bibinfo{author}{\bibfnamefont{C.}~\bibnamefont{Eckberg}},
	\bibinfo{author}{\bibfnamefont{P.}~\bibnamefont{Deng}},
	\bibinfo{author}{\bibfnamefont{S.}~\bibnamefont{Krylyuk}},
	\bibinfo{author}{\bibfnamefont{A.~V.} \bibnamefont{Davydov}},
	\bibnamefont{et~al.}, \bibinfo{journal}{Nature Communications}
	\textbf{\bibinfo{volume}{14}}, \bibinfo{pages}{6691} (\bibinfo{year}{2023}),
	ISSN \bibinfo{issn}{2041-1723}, \bibinfo{note}{publisher: Nature Publishing
		Group}, \urlprefix\url{https://www.nature.com/articles/s41467-023-42447-4}.
	
	\bibitem[{\citenamefont{Le et~al.}(2024)\citenamefont{Le, Pan, Xu, Liu, Wang,
			Lou, Yang, Wang, Yao, Wu et~al.}}]{le_superconducting_2024}
	\bibinfo{author}{\bibfnamefont{T.}~\bibnamefont{Le}},
	\bibinfo{author}{\bibfnamefont{Z.}~\bibnamefont{Pan}},
	\bibinfo{author}{\bibfnamefont{Z.}~\bibnamefont{Xu}},
	\bibinfo{author}{\bibfnamefont{J.}~\bibnamefont{Liu}},
	\bibinfo{author}{\bibfnamefont{J.}~\bibnamefont{Wang}},
	\bibinfo{author}{\bibfnamefont{Z.}~\bibnamefont{Lou}},
	\bibinfo{author}{\bibfnamefont{X.}~\bibnamefont{Yang}},
	\bibinfo{author}{\bibfnamefont{Z.}~\bibnamefont{Wang}},
	\bibinfo{author}{\bibfnamefont{Y.}~\bibnamefont{Yao}},
	\bibinfo{author}{\bibfnamefont{C.}~\bibnamefont{Wu}}, \bibnamefont{et~al.},
	\bibinfo{journal}{Nature} \textbf{\bibinfo{volume}{630}}, \bibinfo{pages}{64}
	(\bibinfo{year}{2024}), ISSN \bibinfo{issn}{1476-4687},
	\bibinfo{note}{publisher: Nature Publishing Group},
	\urlprefix\url{https://www.nature.com/articles/s41586-024-07431-y}.
	
	\bibitem[{\citenamefont{Han et~al.}(2025)\citenamefont{Han, Lu, Hadjri, Shi,
			Wu, Xu, Yao, Cotten, Sharifi~Sedeh, Weldeyesus et~al.}}]{Han2025}
	\bibinfo{author}{\bibfnamefont{T.}~\bibnamefont{Han}},
	\bibinfo{author}{\bibfnamefont{Z.}~\bibnamefont{Lu}},
	\bibinfo{author}{\bibfnamefont{Z.}~\bibnamefont{Hadjri}},
	\bibinfo{author}{\bibfnamefont{L.}~\bibnamefont{Shi}},
	\bibinfo{author}{\bibfnamefont{Z.}~\bibnamefont{Wu}},
	\bibinfo{author}{\bibfnamefont{W.}~\bibnamefont{Xu}},
	\bibinfo{author}{\bibfnamefont{Y.}~\bibnamefont{Yao}},
	\bibinfo{author}{\bibfnamefont{A.~A.} \bibnamefont{Cotten}},
	\bibinfo{author}{\bibfnamefont{O.}~\bibnamefont{Sharifi~Sedeh}},
	\bibinfo{author}{\bibfnamefont{H.}~\bibnamefont{Weldeyesus}},
	\bibnamefont{et~al.}, \bibinfo{journal}{Nature}
	\textbf{\bibinfo{volume}{643}}, \bibinfo{pages}{654} (\bibinfo{year}{2025}),
	ISSN \bibinfo{issn}{1476-4687}.
	
\end{thebibliography}
\end{document}